\documentclass[twocolumn,showpacs,preprintnumbers,amsmath,amssymb]{revtex4-1}
\usepackage{comment} 
\usepackage{graphicx}
\usepackage{dcolumn}
\usepackage{bm}
\usepackage[compact]{titlesec}

\usepackage{xcolor}

\usepackage{hyperref}
\renewcommand{\arraystretch}{1.8}       
\newcommand{\half}{{{\textstyle\frac{1}{2}}}}
\newcommand{\quarter}{{{\textstyle\frac{1}{4}}}}
\newcommand{\be}{\begin{equation}}
\newcommand{\ee}{\end{equation} }
\newcommand{\beqa}{\begin{eqnarray} }
\newcommand{\eeqa}{\end{eqnarray} }
\newcommand{\ba}{\begin{array}}
\newcommand{\ea}{\end{array}}
\newcommand{\bpm}{\begin{pmatrix}}
\newcommand{\epm}{\end{pmatrix}}

\newcommand{\so}{\mathbf{so}}

\newcommand{\Spin}{\mathbf{Spin}}

\newcommand{\dis}{\displaystyle}

\newcommand{\rmd}{{\rm d}}
\newcommand{\rd}{{\rmd}}

\newcommand{\rmD}{{\rm D}}

\newcommand{\ODD}{\mathbf{O}(D,D)}

\newcommand\cE{{\cal E}}

\newcommand\cH{{\cal H}}

\newcommand\cJ{{\cal J}}

\newcommand\cR{{\cal R}}

\def\tx{\tilde{x}}

\def\brn{{\bar{n}}}

\def\brP{\bar{P}}

\newcommand{\Lp}{L_{\varphi}}

\newcommand\vP{{\vec{P}}}

\newcommand\vx{{\vec{x}}}

\newcommand{\rc}{r_{c}}

\newcommand{\trd}{{\bigtriangledown}}

\begin{document}

\title{Identifying Riemannian singularities with regular non-Riemannian geometry}

\author{Kevin Morand}
\email{morand@sogang.ac.kr}

\author{Jeong-Hyuck Park}
\email{park@sogang.ac.kr}
\affiliation{Department of Physics, Sogang University, 35 Baekbeom-ro, Mapo-gu, Seoul 04107,  Korea}

\author{Miok Park}
\email{miokpark76@ibs.re.kr}
\affiliation{{Center for Theoretical Physics of the Universe, Institute for Basic Science, Daejeon 34126, 
Korea}}\email{Valid from August 2021.}
\affiliation{School of Physics, Korea Institute for Advanced Study, Seoul 02455, Korea}

\begin{abstract}
\noindent  
Admitting non-Riemannian geometries, Double Field Theory   extends the notion of spacetime  beyond the Riemannian paradigm.  We   identify    a class of   singular  spacetimes    known in General Relativity with   
 regular non-Riemannian  geometries. The former divergences     merely  correspond to  coordinate singularities of  the  generalised   metric for the latter.   Computed   in  string frame, they  feature an impenetrable non-Riemannian sphere    outside of which    geodesics are complete  with   no singular  deviation.  Approaching  the non-Riemannian points, particles freeze and  strings become (anti-)chiral.
\end{abstract}

                             
\maketitle

\section{Introduction}
Spacetime singularities urge  General Relativity (GR) to evolve. If not in---still elusive---quantum gravity,  the  singularities  in any \textit{Riemannian} metric-based classical  theories of gravity  have several layers:  \textbf{i)} curable,  coordinate singularity of the metric, \textbf{ii)} genuine, curvature  singularity, and  \textbf{iii)}  geodesic incompleteness, as featured {prominently} in  Penrose  theorem~\cite{Penrose:1964wq}.  However,  on the one hand,  \textbf{ii)}  does not necessarily  imply      \textbf{iii)}  (see  \textit{e.g.~}\cite{Olmo:2015bya} for  a recent example)  and,  on the other hand,  one encounters  an   intrinsic  ambiguity in   choosing   a {frame}~\cite{Faraoni:1998qx,Faraoni:2006fx} when  applying these notions  to  scalar-tensor theories,\,\textit{e.g.\,}\cite{Horndeski:1974wa}, since   Weyl transformations  simply do not leave  curvatures and  geodesics  invariant.

From a string theory perspective, the Riemannian metric  $g_{\mu\nu}$  is only a   fraction of  the  massless sector of closed strings   which should further   contain   a skew-symmetric $B$-field and a scalar dilaton $\phi$.     The theory   implies  a   crucial  symmetry, $\ODD$ with   spacetime dimension~$D$, which  transforms the trio $\{g,B,\phi\}$   into one another~\cite{Buscher:1987sk,Buscher:1987qj}. It calls  for   an $\ODD$ singlet, $e^{-2d}=\sqrt{-g}e^{-2\phi}$,  and   an $\ODD$ tensor,
\be
\cH_{AB} = \scalebox{0.9}{\!$ 
\left(\!\ba{cc}
g^{-1} & - g^{-1}B \\
B g^{-1} & g - B g^{-1} B
\ea\!\right)\!=\! \left(\!\!\ba{cc}
1 & 0 \\
B &1
\ea\!\right) \!\!
\left(\!\!\ba{cc}
g^{-1} &\!0 \\
0 &\!g 
\ea\!\!\right) \!\!
\left(\!\!\ba{cc}
1 & -B\\
0 &1
\ea\!\!\right)$}
\,,
\label{RcH}
\ee
which served as  \textit{generalised metric}  in the $\ODD$ manifest  formulations of both  string worldsheet   actions~\cite{Giveon:1988tt,Duff:1989tf,Tseytlin:1990nb,Tseytlin:1990va,Rocek:1991ps,Giveon:1991jj,Hull:2004in,Hull:2006qs,Hull:2006va}  and   target spacetime effective descriptions currently  called  Double Field Theory (DFT)~\cite{Siegel:1993xq,Siegel:1993th,Hull:2009mi,Hull:2009zb,Hohm:2010jy,Hohm:2010pp}.   By taking  
 not  $\{g,B,\phi\}$ but    the $\ODD$ multiplets themselves   as the fundamental variables, DFT opens  a new avenue  beyond the  Riemannian paradigm~\cite{Lee:2013hma,Morand:2017fnv,Cho:2019ofr,Park:2020ixf}.  In this approach, $e^{-2d}$ is   an elementary   scalar density with a unit diffeomorphic weight,  and  $\cH_{AB}$   satisfies   its own   defining properties,
\be
\ba{ll}
\cH_{AB}=\cH_{BA}\,,\qquad&\qquad
\cH_{A}{}^{C}\cH_{B}{}^{D}\cJ_{CD}=\cJ_{AB}\,.
\ea
\label{defcH}
\ee 
Here $\cJ_{AB}{\,=\!\scalebox{0.6}{$\mathbf{\left(\ba{cc}\mathbf{0}&\mathbf{1}
\\\mathbf{1}&\mathbf{0}\ea\right)}$}}$ is the $\ODD$  invariant metric  which, with its inverse,  lowers and raises  the  $\ODD$  indices,  $A,B=1,2,\ldots, 2D$.   Seen as a square matrix, $\cH_{A}{}^{B}$ squares to identity hence is invertible with $\left|\det\cH\right|=1$. 

Remarkably, DFT can be formulated  in terms of any generalised metric that satisfies the defining relations~(\ref{defcH}), and  may have evolved  to  an alternative `pure gravity'  in the sense that both $d$ and $\cH_{AB}$, \textit{i.e.~}the whole closed string massless sector, are taken as the fundamental, gravitational fields.  It is by now fully equipped with  \textit{its own}   Christoffel symbols~$\Gamma_{ABC}$,     scalar/Ricci/Einstein curvatures~\cite{Jeon:2011cn,Park:2015bza}, and   Einstein  equations coupled to extra `matter':  $G_{AB}=8\pi G\,T_{AB}$~\cite{Angus:2018mep}. This single expression   unifies    the equations of motion of  the fundamental variables  $\{d,\cH_{AB}\}$.    Besides,    as solutions to condition~(\ref{defcH}), all  possible  classical  geometries  which DFT is capable of describing   have been  classified by two non-negative integers, $(n,\brn)$~\cite{Morand:2017fnv}. Only  the type $(0,0)$  is   Riemannian,  amounting to   the well-known parametrisation~(\ref{RcH}),  while  others are non-Riemannian in nature:   the  upper left ${D\times D}$   block of the generalised metric  is degenerate with    `nullity', \textit{i.e.~}dimension of the kernel,  ${n+\brn}$, and thus does not yield  an invertible  Riemannian metric.  For a non-Riemannian geometry to be  a consistent string background at the quantum level,      it  turns out     necessary to put   ${n=\brn}$  in the usual  critical dimensions,  ${D=10}$ or $26$~\cite{Park:2020ixf}.  Non-relativistic string~\cite{Gomis:2000bd,Danielsson:2000gi,Gomis:2005pg}, or   torsional/stringy Newton--Cartan  gravities of recent  interest~\cite{Christensen:2013lma,Hartong:2015zia,Harmark:2017rpg,Harmark:2018cdl,Bergshoeff:2018yvt,Bergshoeff:2019pij,Harmark:2019upf,Bergshoeff:2021bmc}    
are of the type $(1,1)$~\cite{Ko:2015rha,Park:2016sbw,Berman:2019izh,Blair:2019qwi,Cho:2019ofr,Gallegos:2020egk,Blair:2020gng,Blair:2021ycc}.      The condition ${n=\brn}$  further   makes  the non-Riemannian geometry   compatible with type II supersymmetric DFT and superstring~\cite{Jeon:2012hp,Park:2016sbw}  {(\textit{c.f.~}\cite{Coimbra:2011nw,Hatsuda:2015cia})}:  for  spin group   $\Spin(t,s){\times\Spin(s,t)}$ with  ${t+s=10}$,    the allowed  range of ${n=\brn}$ is from zero  to  $\min(t,s)$~\cite{Morand:2017fnv}.

Formally, DFT employs    a doubled coordinate system, $x^{A}=(\tx_{\mu},x^{\nu})$, and  subsequently  unifies      diffeomorphisms and $B$-field gauge symmetry {into  `doubled' diffeomorphisms}. Yet,   the   geometry  is not truly doubled:  half of the coordinate dependency should be turned off  \textit{e.g.} by setting   ${\scalebox{1.1}{${\frac{\partial~}{\partial\tx_{\mu}}}$}\equiv0}$, up to $\ODD$ rotations.   This may suggest that  half of the doubled coordinates are actually \textit{gauged}~\cite{Park:2013mpa}.  By gauging  \textit{e.g.} $\tx_{\mu}$  explicitly as $\rmD x^{A}=\big(\rmd\tx_{\mu}-A_{\mu},\rmd x^{\nu}\big)$,   it is possible to define an   $\ODD$-symmetric    proper length~\cite{Park:2017snt} {by means of} the corresponding  pull-back of the   generalised metric $\rmD x^{A}\rmD x^{B}\cH_{AB}$,  and to  further    construct   `doubled-yet-gauged'  actions for  particles~\cite{Ko:2016dxa,Basile:2019pic} and strings~\cite{Lee:2013hma,Park:2016sbw}.  When the generalised metric is $(0,0)$  Riemannian,    all the components of the auxiliary gauge connection~$A_{\mu}$ appear quadratically in the actions. After  integrating  them  out, one recovers the  conventional   particle and string actions. On the other hand, when the background is non-Riemannian,  ${n+\brn}$  components   come out linearly to play the role of Lagrange multipliers. Consequently,  the particle is frozen with identically vanishing  proper velocities  over the non-Riemannian $({n+\brn})$-dimensions. Further,  the string  becomes  chiral over the $n$-directions  and anti-chiral over the $\brn$-directions~\cite{Morand:2017fnv},  as happens   in  the $(1,1)$ non-relativistic  string case~\cite{Gomis:2000bd}.  This also implies that, at the classical level,  chiral strings   get  frozen too~\cite{Park:2020ixf}.

From the Riemannian  perspective, all the non-Riemannian backgrounds  are singular  geometries, as the    would-be Riemannian metric    diverges. Contradistinctly, they are  well-behaved regular  geometries  in the doubled framework.    This motivates us  to revisit    known   singular Riemannian  geometries   and  examine   their non-Riemannian regularity, \textit{c.f.~}earlier related discussions in \cite{Malek:2013sp,Berman:2014hna,Ko:2016dxa,Blair:2016xnn}.   In this Letter,    we report that   a class of  curvature  singularities in GR are    \textit{regular} non-Riemannian geometries  of DFT. They are {at worst}   coordinate singularities  of  DFT.  Moreover, we show  through examples     that the  ordinary  (undoubled)  geodesics defined in their  string frame---which descend  from $\ODD$-symmetric (doubled) geodesics---are \textit{complete} and that   physically measurable tidal forces   do \textit{not}  diverge, in spite of    Riemannian curvature  singularities.   This last statement, although motivated by the DFT perspective,  holds  independently  within the conventional  GR setup.     \vspace{3pt}

\section{Main Idea \& Results}
The Riemann-wise singular  geometries of our interest assume  the following generic form, with $x^{\mu}=(t,y,x^{i})$, \vspace{-5pt}
\be
\ba{cll}
\rd s^{2}&\!=\!&\frac{1}{F}\left(-\rd t^{2} +\rd y^{2}\right)+G_{ij}\rd x^{i}\rd x^{j}\,,\\
B_{(2)}&\!=\!&\pm\frac{1}{F}\,\rd t\wedge \rd y + \frac{1}{2}\,\beta_{\mu\nu}\rd x^{\mu}\wedge\rd x^{\nu} \,,\\
e^{-2\phi}&\!=\!&F \Psi\,.\vspace{-5pt}
\ea
\label{ANSATZ}
\ee
Our non-exhaustive list of examples includes    \textbf{A.} ${D=2}$ black hole by  Witten~\cite{Witten:1991yr},  \textbf{B.} ${D=4}$ spherical solution by Burgess \textit{et al.}~\cite{Burgess:1994kq},  and    \textbf{C.} ${D=10}$ black $5$-brane by Horowitz and Strominger~\cite{Horowitz:1991cd}.  
All  of them  feature  curvature singularities  where $F$ vanishes, while  $G_{ij},\beta_{\mu\nu},\Psi$  remain harmlessly  regular.   
Substituting   (\ref{ANSATZ}) into (\ref{RcH}) yields  the crucial observation (\textit{c.f.~}\cite{Lee:2013hma,Blair:2015eba,Berman:2019izh,Blair:2019qwi} for earlier examples) that the coordinate singularity is absent  in the $\ODD$   fundamental variables: no  negative power of $F$ appears,\vspace{-3pt}
\be
\ba{ll}
e^{-2d}=\Psi\sqrt{G}\,,\qquad&\qquad
\cH=\scalebox{0.8}{$\left(\!\!\ba{cc}
\mathbf{1} & \mathbf{0} \\
\,\bf{\beta}\,&\mathbf{1}
\ea\!\right)$} \mathring{\cH}
\scalebox{0.8}{$\left(\!\!\ba{cc}
\mathbf{1} & \bf{-\beta}\\
\mathbf{0} &\mathbf{1}
\ea\!\!\right)$}\,,
\ea \vspace{-9pt}
\label{cHd}
\ee 
where,  with   Pauli matrices, 
\be
\mathring{\cH}_{AB}=\scalebox{0.8}{$\left(\ba{cccc}-F\sigma_{3}&0&~~\pm\sigma_{1}~~& ~~0~~\\0&G^{-1}&0&0\\\pm\sigma_{1}&0&0&0\\
0&0&0&G\ea\right)$}\,.
\label{cHd2} 
\ee
Clearly, at the points where ${F=0}$,    the  DFT  geometry  is   non-Riemannian, specifically  of  type $(1,1)$, as expected from the underlying  Minkowskian  spin group signature.  

 {We now revisit} the  three aforementioned layers of singularity  from the `doubled' perspective. 

\subsection*{\textbf{i{)\!)}} \textit{Removing the coordinate singularity from  generalised metric through doubled diffeomorphisms.}}\noindent

{
The singular term in $B_{(2)}$ is crucial  in regularising $\cH$. In fact, in the  examples below, this term being pure gauge would not be present when writing the solutions in their simplest form. We shall nevertheless introduce such a singular pure gauge term in order to match the privileged doubled coordinate system~\eqref{ANSATZ}. 
  Namely,  through doubled diffeomorphisms, one can eliminate what should be construed as a \textit{coordinate singularity} of the generalised metric.}
\subsection*{\textbf{ii{)\!)}} \textit{All the $\ODD$-symmetric curvatures are regular.}}\noindent
Once the DFT fundamental variables $\{\cH_{AB},e^{-2d}\}$ are made  free of any singularity,  (and   twice continuously differentiable),  all the $\ODD$-symmetric curvatures  are automatically  regular.  In fact,   all our  examples   satisfy the equations of motion of the NS-NS supergravity, possibly  with an $\ODD$-symmetric  ``cosmological constant'':\vspace{-5pt}
\begin{eqnarray}
&
\scalebox{0.97}{$R+4\Box\phi-4\partial_{\mu}\phi\partial^{\mu}\phi-
\textstyle{\frac{1}{12}}H_{\mu\nu\rho}H^{\mu\nu\rho}-2\Lambda_{\scriptscriptstyle{\rm{DFT}}}=0\,,~~~~$}&\label{NSNS}\\
&\scalebox{0.97}{\!$R_{\mu\nu}+2\trd_{\mu}(\partial_{\nu}\phi)-\quarter H_{\mu\rho\sigma}H_{\nu}{}^{\rho\sigma}
=0\,,\quad
\rd\star\!\left(e^{-2\phi}H_{\scriptscriptstyle{(3)}}\right)=0\,.$}&\nonumber
\end{eqnarray}
This readily  implies that the corresponding non-singular   DFT ansatz~(\ref{cHd}) solves  the DFT   Einstein equations  ${G_{AB}+\cJ_{AB}\,\Lambda_{\scriptscriptstyle{\rm{DFT}}}=0}$ \textit{everywhere}  even at  the non-Riemannian points of ${F=0}$.   It also means  that   all  the $\ODD$-symmetric DFT  curvatures  are    trivially regular~\cite{Berman:2014hna}. Similarly,  while the conventional dilaton~$\phi$ diverges as  ${F\rightarrow 0}$, the DFT dilaton $d$ remains finite, \textit{c.f.~}(\ref{cHd}), hence so does the $\ODD$-symmetric version of  the Fradkin--Tseytlin term~\cite{Fradkin:1984pq}  in  doubled string actions~\cite{Hull:2006va,Fernandez-Melgarejo:2018wpg}.

\subsection*{\textbf{iii{)\!)}}  \textit{Geodesics are complete in string frame:\\~\quad  impenetrable non-Riemannian sphere}}\noindent
The fact that the DFT dilaton $e^{-2d}$ carries a nontrivial diffeomorphic weight prevents it from  coupling  to the $\ODD$-symmetric doubled-yet-gauged particle and string actions~\cite{Ko:2016dxa,Lee:2013hma}, { hence to particle and stringy geodesics, whose equations read respectively}~\cite{Morand:2017fnv,Foot2nd}:
\be
\ba{r}
\textstyle{e\frac{\rd~}{\rd\tau}(e^{-1}\cH_{AB}\rmD_{\tau}x^{B})+2\Gamma_{ABC}(\brP\rmD_{\tau}x)^{B}(P\rmD_{\tau}x)^{C\!}=0\,,}\\
\textstyle{\frac{1}{\sqrt{-h}}}\partial_{\alpha}(\sqrt{-h}\cH_{AB}\rmD^{\alpha}x^{B})+\Gamma_{ABC}(\brP\rmD_{\alpha}x)^{B}(P\rmD^{\alpha}x)^{C\!}=0\,,
\ea
\label{GEO}
\ee
where $e,h_{\alpha\beta}$ are the worldline einbein and worldsheet  metric, while  $P_{AB},\brP_{AB}$ stand for $\half(\cJ_{AB}\pm\cH_{AB})$ {and $\rmD_\tau x$, $\rmD_\alpha x$ are the pull-backs of the doubled-yet-gauged differential.}  This rigidity   naturally settles  the issue of   the  frame ambiguity.  Upon the Riemannian background of (\ref{RcH}), an $\ODD$-symmetric particle   follows  geodesics  defined   in the  \textit{string frame}~\cite{FootPenrose}: after fixing  gauges and solving for the auxiliary connection, (\ref{GEO})  can be shown to reproduce     the standard (undoubled)  expressions, 
\be
\ba{r}
\ddot{x}^{\mu}+\gamma^{\mu}_{\nu\rho}\dot{x}^{\nu}\dot{x}^{\rho}=0~~\Longleftrightarrow~~ 
\frac{\rm d~}{\rm d\lambda}(g_{\mu\nu}\dot{x}^{\nu})-\half \partial_{\mu}g_{\nu\rho}\dot{x}^{\nu}\dot{x}^{\rho}=0\,,\\
\!\scalebox{0.92}{$\partial_{+}(g_{\mu\nu}\partial_{-}x^{\nu})+\partial_{-}(g_{\mu\nu}\partial_{+}x^{\nu}){+
(}H_{\mu\nu\rho}-\partial_{\mu}g_{\nu\rho})\partial_{+}x^{\nu}\partial_{-}x^{\rho}=0\,.$}
\ea
\label{sd}
\ee
Focusing on the ansatz~(\ref{ANSATZ}) and by analogy with \textbf{ii{)\!)}},  the doubled formulation~(\ref{GEO}) in terms of the generalised metric~$\cH_{AB}$ may still suggest that geodesics are regular in the limit $F\rightarrow 0$, despite the fact that its undoubled counterpart~(\ref{sd}) involves the singular Riemannian metric~$g_{\mu\nu}$. 
Indeed,  
as  we show below for     the aforementioned  three   examples,   both null and time-like geodesics are  complete (at least)  in the region ${F>0}$. The   non-Riemannian points of  ${F=0}$  form a    sphere for ${D>2}$ (or  hyperbola for ${D=2}$)  inside of which (${F<0}$) no time-like nor null geodesics  can pene\-trate.  Time-like and non-radial null geodesics cannot  even come close to  the sphere  from outside (${F>0}$). Only radial null ones may approach  the sphere with identically vanishing proper  velocities  taking infinite affine parameter.   Besides,  undoubled  strings---as  an alternative probe of the ``singular"  geometries~\cite{Duff:1991pe,Duff:1994an}---become  (anti-)chiral: with  light-cone coordinates $y^{\pm}=y\pm t$  and also  $\partial_{\pm\!}=\!\frac{\partial~}{\partial\sigma^{\pm}}$ on the worldsheet, as already used in (\ref{sd}),   we get $\partial_{-}y^{+}=0=\partial_{+}y^{-}$ (or $\partial_{+}y^{+}=0=\partial_{-}y^{-}$)  in the limit  ${F\rightarrow 0}$.   Thus,   the undoubled, or conventional,  particle  geodesics and string propagations  agree with  the   non-Riemannian  (freezing/chiral/anti-chiral) behaviors  predicted from  the  previous  doubled-yet-gauged sigma model approach~\cite{Lee:2013hma,Morand:2017fnv} which  relied on   the  auxiliary gauge connection as Lagrange multiplier (\textit{c.f.} Introduction). 

\subsection*{\mbox{\!\textbf{ii{)\!)}} \textbf{\&} \textbf{iii{)\!) again:}} geodesic deviations  are also regular.}}
In DFT   there is no $\ODD$-symmetric completion of the  Riemann curvature~\cite{Jeon:2010rw,Jeon:2011cn,Hohm:2011si}. Hence,  the criterion  of `$\ODD$-symmetric curvature singularity' may  appear   unbalanced or somewhat  unfair compared to GR.    
As a step toward restoring the balance as well as focusing on genuine physical quantities,
     we further analyse    the geodesic deviation  and the  `tidal force'  therein (again in string frame), \vspace{-2pt}
\be
\scalebox{1.1}{$
\frac{\rm{D}^{2}\xi^{\mu}}{\rd\lambda^{2}}$}=R^{\mu}{}_{\nu\rho\sigma}\dot{x}^{\nu}\dot{x}^{\rho}\xi^{\sigma}\,.
\label{geodev}
\ee  
{As the geodesics are   complete and smooth  for ${F\geq0}$, their deviations~$\xi^{\mu}$ should be regular.  Then, although  the Riemann curvature itself diverges,  its contraction with the vanishing  velocities $R^{\mu}{}_{\nu\rho\sigma}\dot{x}^{\nu}\dot{x}^{\rho}$  as well as    the square norm $\big|\frac{\rm{D}^{2}\xi}{\rd\lambda^{2}}\big|^{2}=g_{\mu\nu}\frac{\rm{D}^{2}\xi^{\mu}}{\rd\lambda^{2}}\frac{\rm{D}^{2}\xi^{\nu}}{\rd\lambda^{2}}$  can  be   finite,  \textit{thus preventing the physically measurable quantities from being singular}. 
This further property will be checked to hold in the following section for the  examples   \textbf{A, B} and \textbf{C}. \vspace{3pt}

\section{Examples: geodesics and string propagation}
We now take a closer look at   each   example~\cite{Witten:1991yr,Burgess:1994kq,Horowitz:1991cd}  to   analyse      generic    geodesics~(\ref{sd}) 
supplemented by  the Hamiltonian or Virasoro constraints,\vspace{-2pt}
\be
\ba{ll}
g_{\mu\nu}\dot{x}^{\mu}\dot{x}^{\nu}=\cE\,,\qquad&\qquad
\partial_{+}x^{\mu}\partial_{+}x^{\nu}g_{\mu\nu}=0=
\partial_{-}x^{\mu}\partial_{-}x^{\nu}g_{\mu\nu}\,.\vspace{-3pt}
\ea
\label{Vira}
\ee
Hereafter, $\cE$ is to be either $-1$ (time-like) or $0$ (null). 

\subsection{${D=2}$   black hole: non-Riemannian hyperbola}
The   ${D=2}$   geometry \`{a} la   Witten~\cite{Witten:1991yr} is characterised  by 
$\rd s^{2}=\rd y^{+}\rd y^{-}/F$ with ${F=-1+y^{+}y^{-}/l^{2}=\frac{F}{|F|}e^{-2\phi}}$.  
 It solves (\ref{NSNS}) when  ${\Lambda_{\scriptscriptstyle{\rm{DFT}}}=-\frac{2}{l^{2}}}$.  From the on-shell value of the Ricci scalar ${R=-\frac{4}{l^{2}F}}$,  the hyperbola ${y^{+}y^{-}=l^{2}}$    corresponds to     a   curvature singularity. We stress that though the $H$-flux is trivial in two dimensions, the $B$-field  still  plays a crucial role    in making  the generalised metric free of  singularity~(\ref{cHd2}).

Since the metric is invariant under  scaling $\delta y^{\pm}=\pm y^{\pm}$,  there are two constants of motion for every geodesic, \vspace{-3pt}
\be
\ba{ll}
L=(y^{-}\dot{y}^{+}-y^{+}\dot{y}^{-})/F\,,\quad
&\quad\cE=\dot{y}^{+}\dot{y}^{-}/F\,,
\ea
\label{LF}
\ee
which give, setting  ${\omega\equiv{l\dot{y}^{+}}/{y^{+}}}$,  
\[
\ba{c}
\scalebox{1}{$\textstyle{
\!{l\frac{\rd F}{\rd\lambda}}\!=\!\pm\sqrt{\!F\!\left[4\cE\!+\!\left({\frac{L^{2}}{l^{2}}\!+4\cE}\right)\! F\right]}\,,~\,
{l^{2}{\frac{\rd^{2}F}{\rd\lambda^{2}}}}\!=\!\left(\!\frac{L^{2}}{l^{2}}\!+\!4\cE\right)\! F{+2\cE}\,,}$}\\
\textstyle{\frac{L^{2}}{l^{2}}+4\cE=\left(\omega{-\frac{\cE}{\omega}}\right)^{2\!}+\left(\frac{\omega}{F}\right)^{2\!}+\frac{2(\omega^{2}{-\cE})}{F}\,.}
\ea
\]
It follows  that   time-like geodesics (${\cE=-1}$) starting from the region ${F>0}$  will  never reach the non-Riemannian hyperbola,  satisfying  ${F\geq \frac{4}{(L/l)^{2}-4}>0}$.   Null ones (${\cE=0}$)  may approach  only at past or future infinity as  ${F=e^{\pm L\lambda/l^{2}}F_{0}}$, since the most  general   null geodesics are,  from (\ref{LF})  with initial values $y^{\pm}_{0}$ and $F_{0}$ at ${\lambda=0}$,    either
\be 
\left(\ba{c}y^{+}(\lambda)\\y^{-}(\lambda)\ea\right)=
\left(\ba{c}
l^{2}/y_{0}^{-}+\big(y_{0}^{+}-l^{2}/y^{-}_{0}\big) e^{L\lambda/l^{2}}\\
y_{0}^{-}\ea\right)\,,
\label{D2SOL}
\ee
or  $y^{+}(\lambda)=y_{0}^{+}$, $y^{-}(\lambda)= l^{2}/y_{0}^{+}+\big(y_{0}^{-}-l^{2}/y^{+}_{0}\big)e^{-L\lambda/l^{2}}$. 
For these two solutions,  the only nonvanishing  components in (\ref{geodev})    are  respectively 
$R^{+}{}_{\nu\rho-}\dot{x}^{\nu}\dot{x}^{\rho}=-(L/y_{0}^{-}l)^{2}$ and 
$R^{-}{}_{\nu\rho+}\dot{x}^{\nu}\dot{x}^{\rho}=-(L/y_{0}^{+}l)^{2}$, which are all finite with ${\big|\frac{\rm{D}^{2}\xi}{\rd\lambda^{2}}\big|^{2}=0}$. In fact,  the   variations of the general solutions~(\ref{D2SOL}) by  the free parameters, $y_{0}^{\pm}$,  also  lead to  finite  deviation vectors, $\xi^{\mu}$. We conclude that    the space ${F>0}$  is  geodesically  complete with no singular deviation.   We turn to the  string dynamics~(\ref{sd}), (\ref{Vira})  which read
\[
\ba{rr}
\partial_{+}\partial_{-}y^{+}-\partial_{+}y^{+}\partial_{-}y^{+}\frac{\partial\ln F}{\partial y^{+}}=0\,,\qquad&\qquad\partial_{+}y^{+}\partial_{+}y^{-}=0\,,\\
\partial_{+}\partial_{-}y^{-}-\partial_{+}y^{-}\partial_{-}y^{-}\frac{\partial\ln F}{\partial y^{-}}=0\,,\qquad&\qquad\partial_{-}y^{+}\partial_{-}y^{-}=0\,.
\ea
\]
The second relation gives either ${\partial_{+}y^{+}=0}$ or ${\partial_{+}y^{-}=0}$. If ${\partial_{+}y^{+}=0}$, the third implies ${\partial_{+}(\frac{1}{F}\partial_{-}y^{-})=0}$ such that  ${\partial_{-}y^{-}=F{\mathbf{f}(\sigma^{-})}}$ for some one-variable function $\mathbf{f}(\sigma^{-})$.  Thus, $\partial_{-}y^{-}$ vanishes when ${F=0}$. On the other hand, if  ${\partial_{+}y^{-}=0}$, the first implies ${\partial_{+}(\frac{1}{F}\partial_{-}y^{+})=0}$ and  ${\partial_{-}y^{+}=F\mathbf{f}(\sigma^{-})}$. Thus,   $\partial_{-}y^{+}$ vanishes  when ${F=0}$.  Similar analysis holds for the last relation. { We conclude that,} in any case, one of $\{y^{+},y^{-}\}$ is chiral and the other is anti-chiral on  the non-Riemannian hyperbola.

\subsection{${D=4}$ two-parameter family of spherical solution}
Our  ${D=4}$ example is a spherical solution from \cite{Burgess:1994kq},  \vspace{-2pt}
\[
\ba{l}
\rd s^{2}=\frac{1}{F(r)}\left(-\rd t^{2} +\rd r^{2}\right)+\cR(r)^2\left(\rd\vartheta^{2}+\sin^{2\!}\vartheta\,\rd\varphi^{2}\right)\,,\\
B_{(2)}=\pm\frac{1}{F(r)}\rd t\wedge \rd r+h\cos\vartheta\,\rd t\wedge\rd \varphi \,,
\ea \vspace{-2pt}
\]
where   $F(r), \cR(r)^{2}$ are,   with  two free parameters, $b,h$~\cite{Ko:2016dxa} {(see also \cite{Angus:2018mep} for their physical interpretations)},
\be
\ba{l}
\frac{1}{F(r)}=\frac{1+\sqrt{1-h^{2}/b^{2}}}{2}\left(\frac{r}{r+b}\right)+
\frac{1-\sqrt{1-h^{2}/b^{2}}}{2}\left(\frac{r+b}{r}\right)\,,\\
\cR(r)^{2}=\big(r+\frac{1}{2}b-\frac{1}{2}\sqrt{b^{2}-h^{2}}\big)^{2}+\frac{1}{4}h^{2}=\frac{r(r+b)}{F(r)}\,.
\ea
\ee
We require $0<|h|\leq b$, such that  the only  singular source  in  the metric is at ${r=0}$ as   $\lim_{r\rightarrow 0}F(r)=0$. The  Ricci scalar diverges at ${r=0}$ as $R\simeq -\frac{2}{\left(b-\sqrt{b^{2}-h^{2}}\right)\scalebox{1}{$r$}\,}$.   Nevertheless,  with  non-singular   $\cR(0)^{-2}$ due to  ${h\neq0}$,   the  generalised metric $\cH_{AB}$,  dilaton   $e^{-2d}=\cR(r)^{2}\sin\vartheta$, and   DFT curvatures   are all   finitely regular.  The non-Riemannian points of ${F=0}$ form a  $2$-sphere  with  nontrivial  proper area  $4\pi\cR(0)^{2}=2\pi b(b-\sqrt{b^{2}-h^{2}})$.

For geodesics, without loss of generality, we put ${\vartheta=\frac{\pi}{2}}$. The conserved  energy $E$ and   angular momentum $\Lp$ set \vspace{-3pt}
\be
\ba{ll}
\dot{t}=EF(r)\,,\qquad&\qquad \dot{\varphi}=\Lp\cR(r)^{-2}\,.
\ea\vspace{-2pt}
\label{ELp}
\ee
In particular,  $\dot{t}$ vanishes on the non-Riemannian sphere. The remaining radial motion reads 
\[
\ba{ll}
0=\dot{r}^{2}+V(r)\,,\quad&\quad
V(r)=\left[-\cE-E^{2}F(r)+\frac{\Lp^{2}}{\cR(r)^{2}}\right]\!F(r)\,.
\ea
\]
From ${\lim_{r\rightarrow 0}V=0}$,  it follows that $\dot{r}$ also gets   trivial at ${r=0}$.  In fact, since $\lim_{r\rightarrow 0} V^{\prime}(r)=\frac{4\Lp^{2}}{b(b-\sqrt{b^{2}-h^{2}})^{2}}-\frac{2\cE}{b-\sqrt{b^{2}-h^{2}}}$ is  positive    for either ${\cE=-1}$, or ${\cE=0}$ with ${\Lp\neq0}$, the corresponding potential is positive close to  ${r=0^{+}}$. Thus,   both the   time-like and the  non-radial null geodesics   cannot  come close to the  sphere from outside. Only the  radial null ones with ${\cE=0=\Lp}$ may do, but it takes  infinite affine parameter as the integral of $\rd\lambda= (EF)^{-1} \rd r $ diverges logarithmically. These results   can be all  attributed to  the  \textit{repulsive}  gravitational force  around  the non-Riemannian sphere, while   it is attractive for large $r\sim\cR$    as  $g_{tt}\sim -1+\frac{b\sqrt{1-h^{2}/b^{2}}}{\cR}$. 

For the radial null geodesics  $\dot{t}=EF=|\dot{r}|$, we also confirm  that   the  deviation~(\ref{geodev}) is  finitely regular:   the only nontrivial  components of  $R^{\mu}{}_{\nu\rho\sigma}\dot{x}^{\nu}\dot{x}^{\rho}$ at ${r=0}$ take  the values  $ \frac{\pm 2E^{2}}{(b-\sqrt{b^{2}-h^{2}})^{2}}$ for  $\mu,\sigma$ being $t$ or $r$,  with ${\big|\frac{\rm{D}^{2}\xi}{\rd\lambda^{2}}\big|^{2}=0}$.  

In the limit, ${r\rightarrow0}$ hence ${F\rightarrow 0}$,  the string dynamics~(\ref{sd}), (\ref{Vira}) implies, with $y^{\pm}=r\pm t$,
\be\vspace{-3pt}
\ba{ll}
\partial_{+}y^{+}\partial_{-}y^{+}F^{\prime}(0)=0\,,\qquad&\qquad\partial_{+}y^{+}\partial_{+}y^{-}=0\,,\\
\partial_{+}y^{-}\partial_{-}y^{-}F^{\prime}(0)=0\,,\qquad&\qquad\partial_{-}y^{+}\partial_{-}y^{-}=0\,,
\ea
\label{yyyy}\vspace{-2pt}
\ee
where $F^{\prime}(0)={\lim_{r\rightarrow 0}}F^{\prime}(r)=\textstyle{\frac{2}{b-\sqrt{b^{2}-h^{2}}}}$ is nonvanishing.  This confirms that  one of $\{y^{+},y^{-}\}$ is chiral while  the other is anti-chiral  on the non-Riemannian  sphere. 

\subsection{${D=10}$ black ${5}$-brane}
One  particular  black $5$-brane geometry  from \cite{Horowitz:1991cd} reads 
\[\vspace{-2pt}
\ba{ll}
\!\rd s^{2}=\frac{-\rd t^{2}+\rd r^{2}}{F(r)}+r^{2}\rd\Omega_{3}^{2}+\rd\vec{x}^{\,2}
\,,\quad&~\, \scalebox{0.95}{$F=1{-({\rc/r})^{2}}=e^{-2\phi}$}\,.
\ea
\]  
The  Ricci scalar  diverges  both at ${r=0}$ and  ${r=\rc}$ as $R=-\frac{4\rc^{4}}{r^{4}(r^{2}-\rc^{2})}$. Though the $H$-flux is trivial, a  pure gauge $B$-field should be introduced, as prescribed in  (\ref{ANSATZ}).   The generalised metric~(\ref{cHd2})  is then  non-Riemannian regular  on  the $3$-sphere of the radius ${r=\rc}$,  but  still singular at ${r=0}$.   We shall see soon  that the  non-Riemannian  sphere forms the  boundary of a geodesically complete   space of  ${F>0}$ which excludes the dangerous  point ${r=0}$. 

The geodesic analysis is similar to  example \textbf{B}  and  reduces to $\dot{t}=EF$  and $\dot{r}^{2}+V(r)=0$ with a  potential involving  non-negative  constants,  $L_{\Omega}^{2}$ (total angular momentum) and   $\vP^{2}$ (extra momentum),   
\[\vspace{-2pt}
V(r)=\left[\scalebox{1}{$-\cE-E^{2}F(r)+\frac{L_{\Omega}^{2}}{r^{2}}+\vP^{2}$}\right]\!F(r)\,.
\]
Since  $V^{\prime}(\rc)=2(-\cE+L_{\Omega}^{2}/\rc^{2}+\vP^{2})/\rc$ is positive for  either ${\cE=-1}$ or ${\cE=0}$ with $L_{\Omega}^{2}/\rc^{2}+\vP^{2}\neq0$,  and  ${V(\rc)}$ vanishes,   the corresponding potential is positive close to ${r=\rc^{+}}$. Thus, time-like and  generic null geodesics  cannot  reach the non-Riemannian sphere.  Only the radial  null ones having  $\dot{t}=EF=|\dot{r}|$ and $L_{\Omega}^{2}=0=\vP^{2}$  can do,  albeit  taking infinite affine parameter     with vanishing proper velocities.  Besides, the  deviation~(\ref{geodev})    is  regular:    the only nontrivial values of    $R^{\mu}{}_{\nu\rho\sigma}\dot{x}^{\nu}\dot{x}^{\rho}$ at ${r=\rc}$  are    $\pm 2E^{2}/\rc^{2}$, for   $\mu,\sigma$  being $t$ or $r$, with ${\big|\frac{\rm{D}^{2}\xi}{\rd\lambda^{2}}\big|^{2}=0}$.

In the limit ${r\rightarrow\rc}$ hence ${F\rightarrow 0}$, the  chirality relations  of  the previous example~(\ref{yyyy})   still hold  after replacing $F^{\prime}(0)$ by $F^{\prime}(\rc)=2/\rc$.

\section*{Discussion}
We have shown that the curvature singularities featured in a large class of GR spacetimes \eqref{ANSATZ} are   mere artifacts of Riemannian geometry. In particular, we have noted the remarkable fact that physically measurable {tidal forces} do not diverge.  {Further examples  include  ${D=10}$  superstring~\cite{Dabholkar:1990yf,Berkeley:2014nza} with negative tension~\cite{Dijkgraaf:2016lym,Arvanitakis:2016zes,Blair:2016xnn,Berman:2019izh,Blair:2019qwi}.  The  corresponding generalised metric was  shown to be regular in \cite{Blair:2016xnn}.}

 Converted to the Einstein frame,  $g^{\rm{\bf E}}_{\mu\nu}= e^{4\phi/(2-D)}g_{\mu\nu}$,    the   ${D>2}$  examples feature  more severe curvature singularities   and become geodesically incomplete~\textit{e.g.~}\cite{Kang:2019owv}, with \textit{singular}  deviations.  ${D=2}$ DFT is essentially   Jackiw--Teitelboim gravity~\cite{Teitelboim:1983ux,Jackiw:1984je}  as   $g^{\rm{\bf JT}}_{\mu\nu}= e^{-2\phi}g_{\mu\nu}$.     Witten's  solution  is then mapped to  flat spacetime.

 Constant non-Riemannian backgrounds  were shown in \cite{Blair:2020gng}  to admit infinite-dimensional isometries. All of them might  be realised  as the asymptotic symmetries of the   non-Riemannian spheres with  large radius  (as ${F^{\prime}\rightarrow 0}$). The associated,  infinitely many,  conserved charges   then might  store all  the information of the in-going  and  freezing  null radial geodesics.  We call for further studies.

\hfill
\vspace{-0.3cm}

\noindent\textbf{\textit{Acknowledgments.}}
{We  wish to thank  Gungwon Kang, Nakwoo Kim, Wontae Kim, and Dong-han Yeom for useful discussions {as well as anonymous referees for helpful suggestions.}   This work is supported by Basic Science Research Program through the National Research Foundation of Korea (NRF) through the grants, NRF-2016R1D1A1B01015196, NRF-2018H1D3A1A01030137 (Brain Pool Program),   NRF-2020R1A6A1A03047877 (Center for Quantum Space Time), and NRF-2021R1C1C1005037.  MP was  further supported by a KIAS Individual Grant (PG062002) at Korea Institute for Advanced Study and  by the Institute for Basic Science (Grant No. IBS-R018-Y1).}


\appendix
\renewcommand\thesection{SM}
\setlength{\jot}{9pt}                 
\renewcommand{\arraystretch}{1.8} 
\begin{center}
	\large\textbf{Supplemental Material}
\end{center}

\section*{${D=10}$ fundamental string with negative tension\label{SM}}

We now expound on the aforementioned example of the  fundamental string~\cite{Dabholkar:1990yf},
\[
\ba{ll}
\rd s^{2}=\frac{\rd y^{+}\rd y^{-}}{F(r)}+\rd\vec{x}^{\,2}\,,\quad&\quad B_{(2)}=\frac{\rd y^{+}\wedge\rd y^{-}}{2F(r)}\,,\vspace{-2pt}
\ea
\]
where ${F=1+\frac{Q}{r^{6}}=e^{-2\phi}}$ is a harmonic function and ${r^{2}=\vec{x}{\cdot\vec{x}}}$. 
The Ricci scalar diverges at ${r=0}$ as  $R=-\frac{126 Q^{2}}{r^{2}(Q+r^{6})^{2}}$, and further at $r=|Q|^{1/6}$ provided that the tension is negative, ${Q<0}$. The  corresponding generalised metric---identified  as a  doubled null wave in \cite{Berkeley:2014nza}---is then    non-Riemannian  regular  at $r=|Q|^{1/6}$~\cite{Berman:2019izh,Blair:2019qwi} but still singular at ${r=0}$.  Similarly to  examples  \textbf{B} and \textbf{C},  the non-Riemannian  $7$-sphere of  radius ${r=|Q|^{1/6}}$  forms the  boundary of the  geodesically complete   space    ${F\geq0}$, which thus excludes    the dangerous  point   ${r=0}$.

The longitudinal   isometries give two constants of motion  $P_{\pm}$ and set   ${\dot{y}^{\pm}=P_{\pm}F}$ which  clearly  vanish when ${F=0}$.  Moreover, time-like  and also non-radial (${L^{2}>0}$)  null geodesics  are  incompatible with ${F=0}$,  due to the fact that  ${|\dot{\vec{x}}|^{2}=\dot{r}^{2}+\frac{L^{2}}{r^{2}}=\cE-P_{+}P_{-}F}$ is  non-negative. Hence ${P_{+}P_{-}\leq 0}$ and the acceleration   $\ddot{\vec{x}}=-3 |Q|P_{+}P_{-}\vec{x}/r^{8}$  is repulsive.  Radial (${L^{2}=0}$) null ones with ${P_{+}P_{-}<0}$ may reach the  sphere only  to bounce back within finite affine parameter. On the contrary, null ones with ${P_{+}P_{-}=0}$  have ${\dot{\vec{x}}=0}$  and thus   $r$  is fixed for ever. In particular,  when ${r=|Q|^{1/6}}$,   from  ${\dot{x}^{\mu}=0}$,   $x^{\mu}$ should be (trivially)   constant.   Anyhow,  time-like and null geodesics are all complete.   We  turn to the deviations  of the  bouncing  null geodesics,\vspace{-5pt}
\be
\ba{l}
y^{\pm}(\lambda)\,\simeq\, y^{\pm}_{0}-3|Q|^{-\frac{1}{3}}P_{\pm}P_{+}P_{-}\lambda^{3}\,,\\
\vx(\lambda)\,\simeq\, \hat{n}_{0}\!\left(|Q|^{\frac{1}{6}}-\frac{3}{2}|Q|^{-\frac{1}{6}}P_{+}P_{-}\lambda^{2}\right)\,.
\ea\vspace{-2pt}
\label{bounce}
\ee
This  expansion  around the bouncing  at ${\lambda=0}$   comes from $F\simeq 6(r|Q|^{-1/6}-1)$, 
and  features all possible constants of motion:  $P_{\pm},y^{\pm}_{0}$ and a unit vector $\hat{n}_{0}$.  Varying each of them,    we acquire  a deviation vector $\xi^{\mu}$. Contrarily to the previous examples, some of  $R^{\mu}{}_{\nu\rho\sigma}\dot{x}^{\nu}\dot{x}^{\rho}$'s  exhibit  singularities   as severe as  ${F^{-3/2}\propto\lambda^{-3}}$ for ${\sigma=\pm}$, or  ${F^{-1}\propto \lambda^{-2}}$ for transverse $\sigma$.  Nonetheless,   the full expression of the tidal force  $R^{\mu}{}_{\nu\rho\sigma}\dot{x}^{\nu}\dot{x}^{\rho}\xi^{\sigma}$  is finite, once contracted with the   deviation vectors of (\ref{bounce}) induced by   $\delta{(P_{+}/P_{-})=0}$ (dilation, ${\xi^{\mu}=\lambda\dot{x}^{\mu}}$),     $\delta{(P_{+}P_{-})=0}$ $(\so(1,1)$ Lorentz),   ${\delta\hat{n}_{0}\cdot\vx=0}$ $(\so(8)$ rotation),  with the exception of   $\delta y_{0}^{\pm}$ (translation).    Translational $\xi^{\mu}$'s are   just constant. Being   longitudinal,  their proper length gets singular, and subsequently so does the tidal force acting on them. Lastly, the string propagation~(\ref{sd}) gives  \vspace{-2pt}
\be
\partial_{+}(F^{-1}\partial_{-}y^{+})=0=\partial_{-}(F^{-1}\partial_{+}y^{-})\,,
\ee
which yields  ${\partial_{-}y^{+}=F{\mathbf{f_{\scriptscriptstyle{-}}}(\sigma^{-})}}$ and  ${\partial_{+}y^{-}=F{\mathbf{f_{\scriptscriptstyle{+}}}(\sigma^{+})}}$, naturally  implying the (anti-)chirality at ${r=|Q|^{1/6}}$.  


\begin{thebibliography}{99}
\bibitem{FootPenrose}
Since the null energy condition  can be broken by the term  $2\trd_{\mu}(\partial_{\nu}\phi)$  in  (\ref{NSNS}), Penrose theorem can be evaded   in the  string frame, thus allowing  for geodesically complete spaces.
\end{thebibliography}
\end{document}